\documentclass[preprint,showpacs,preprintnumbers,amsmath,amssymb]{revtex4}
\usepackage{booktabs}
\usepackage{mathrsfs}
\usepackage{epsfig}
\usepackage{graphicx}
\usepackage{dcolumn}
\usepackage{bm}
\usepackage{amsmath}

\let\jnfont=\rm
\def\NPB#1,{{\jnfont Nucl.\ Phys.\ B }{\bf #1},}
\def\PLB#1,{{\jnfont Phys.\ Lett.\ B }{\bf #1},}
\def\EPJC#1,{{\jnfont Eur.\ Phys.\ J.\ C }{\bf #1},}
\def\PRD#1,{{\jnfont Phys.\ Rev.\ D }{\bf #1},}
\def\PRL#1,{{\jnfont Phys.\ Rev.\ Lett.\ }{\bf #1},}
\def\MPLA#1,{{\jnfont Mod.\ Phys.\ Lett.\ A }{\bf #1},}
\def\JPG#1,{{\jnfont J.\ Phys.\ G}{\bf #1},}
\def\CTP#1,{{\jnfont Commun.\ Theor.\ Phys.\ }{\bf #1},}
\def\ZPC#1,{{\jnfont Z.\ Phys.\ C }{\bf #1},}
\def\JHEP#1,{{\jnfont JHEP \ }{\bf #1},}
\def\Rv{\not{\hbox{\kern-1pt $R$}}}
\def\p{\not{\hbox{\kern-3pt $p$}}}
\def\gsim{\raise0.3ex\hbox{$>$\kern-0.75em\raise-1.1ex\hbox{$\sim$}}}
\begin{document}

\title{Rare $Z$-decay into light CP-odd Higgs bosons: a comparative study in
       different new physics models}

\author{Junjie Cao$^1$, Zhaoxia Heng$^2$, Jin Min Yang$^2$}

\address{
 $^1$  Department of Physics,
       Henan Normal University, Xinxiang 453007, China \\
 $^2$ Key Laboratory of Frontiers in Theoretical Physics,
      Institute of Theoretical Physics, Academia Sinica, Beijing 100190, China
\vspace*{1cm}}

\begin{abstract}
Various new physics models predict a light CP-odd Higgs boson
(labeled as $a$) and open up new decay modes for $Z$-boson, such as
$Z \to \bar{f} f a$, $Z\to a\gamma$ and $Z\to aaa$, which could be
explored at the GigaZ option of the ILC. In this work we investigate
these rare decays in several new physics models, namely the type-II
two Higgs doublet model (type-II 2HDM), the lepton-specific two
Higgs doublet model (L2HDM), the nearly minimal supersymetric
standard model (nMSSM) and the next-to-minimal supersymmetric
standard model (NMSSM). We find that in the parameter space allowed
by current experiments, the branching ratios can reach $10^{-4}$ for
$Z \to \bar{f} f a $ ($f=b,\tau$), $10^{-9}$ for $Z\to a\gamma$ and
$10^{-3}$ for $Z\to aaa$, which implies that the decays $Z \to
\bar{f} f a$ and $Z \to a a a$ may be accessible at the GigaZ
option. Moreover, since different models predict different patterns
of the branching ratios, the measurement of these rare decays at the
GigaZ may be utilized to distinguish the models.

\end{abstract}
\maketitle

\section{INTRODUCTION}
The LEP experiments at the resonance of $Z$-boson have tested the
Standard Model (SM) at quantum level, measuring the $Z$-decay into
fermion pairs with an accuracy of one part in ten thousands. The
good agreement of the LEP data with the SM predictions have severely
constrained the behavior of new physics at the $Z$-pole. Taking
these achievements into account one can imagine that the physics of
$Z$-boson will again play the central role in the frontier of
particle physics if the next generation $Z$ factory comes true with
the generated $Z$ events several orders of magnitude higher than
that of the LEP. This factory can be realized in the GigaZ option of
the International Linear Collider (ILC)\cite{gigaz}. The ILC is a
proposed electron-positron collider with tunable energy ranging from
$400 {\rm GeV}$ to $500 {\rm GeV}$ and polarized beams in its first
phase, and the GigaZ option corresponds to its operation on top of
the  resonance of $Z$ boson by adding a bypass to its main beam
line. Given the high luminosity, ${\cal{L}} = 7 \times 10^{33}
cm^{-2} s^{-1} $, and the cross section at the resonance of $Z$
boson, $\sigma_Z \simeq 30 {\rm nb}$, about $2 \times 10^9$ $Z$
events can be generated in an operational year of $10^7 s$ of GigaZ,
which implies that the expected sensitivity to the branching ratio
of $Z$-decay can be improved from $10^{-5}$ at the LEP to $10^{-8}$
at the GigaZ\cite{gigaz}. In light of this, the $Z$-boson
properties, especially its exotic or rare decays which are widely
believed to be sensitive to new physics, should be investigated
comprehensively to evaluate their potential in probing new physics.

Among the rare $Z$-decays, the flavor changing (FC) processes
were most extensively studied to explore the flavor texture in new
physics \cite{Z-review}, and it was found that, although these
processes are severely suppressed in the SM, their branching ratios
in new physics models can be greatly enhanced to $10^{-8}$ for
lepton flavor violation decays \cite{Z-decay-susy} and $10^{-6}$ for
quark flavor violation decays \cite{Z-decay-qq}. Besides the FC
processes, the $Z$-decay into light Higgs boson(s) is another type
of rare process that was widely studied, e.g. the decay $Z \to
\bar{f} f a $ ($f=b, \tau$) with the particle $a$ denoting a light
Higgs boson was studied in \cite{ffa}, the decay $Z \to a \gamma $
was studied in the two Higgs doublet model (2HDM)\cite{agamma-2HDM}
and the minimal supersymmetric standard model
(MSSM)\cite{agamma-MSSM}, and the decay $Z\to aaa$ was studied in a
model independent way \cite{aaa1}, in 2HDM\cite{aaa2,yuan} and also
in MSSM\cite{aaa3}. These studies indicate that, in contrast with
the kinematic forbidden of these decays in the SM, the rates of
these decays can be as large as $10^{-5}$ in new physics models,
which lie within the expected sensitivity of the GigaZ.  In this
work, we extend the previous studies of these decays to some new
models and investigate these decays altogether. We are motivated by
some recent studies on the singlet extension of the MSSM, such as
the next-to-minimal supersymmetric standard model (NMSSM)
\cite{NMSSM,NMSSM-1} and the nearly minimal supersymmetric standard
model (nMSSM) \cite{nMSSM}, where a light CP-odd Higgs boson $a$
with singlet-dominant component may naturally arise from the
spontaneous breaking of some approximate global symmetry like
$U_R(1)$ or Peccei-Quuin symmetry
\cite{light-a-NMSSM,light-a-NMSSM-1,Gunion}. These non-minimal
supersymmetric models can not only avoid the $\mu$-problem, but also
alleviate the little hierarchy by having such a light Higgs boson
$a$ \cite{little-hierachy}. We are also motivated
by that, with the latest experiments, the properties of the light
Higgs boson are more stringently constrained than before. So it is
worth updating the previous studies.

So far there is no model-independent lower bound on the lightest
Higgs boson mass. In the SM, it must be heavier than $114$ GeV,
obtained from the null observation of the Higgs boson at LEP
experiments. However, due to the more complex structure of the Higgs
sector in the extensions of the SM, this lower bound can be
significantly relaxed according to recent studies, e.g., for the
CP-odd Higgs boson $a$ we have $m_a\gsim 1$ GeV in the nMSSM
\cite{nMSSM-cao}, $m_a\gsim 0.21$ GeV in the NMSSM
\cite{light-a-NMSSM-1}, and $m_a\gsim 7$ GeV in the lepton-specific
2HDM (L2HDM) \cite{l2hdm-cao}. With such a light CP-odd Higgs boson,
the Z-decay into one or more $a$ is open up. Noting that the decay
$Z\to aa$ is forbidden due to Bose symmetry, we in this work study
the rare $Z$-decays $Z \to \bar{f} f a $ ($f=b, \tau$), $Z\to
a\gamma$ and $Z\to aaa$  in a comparative way for four models,
namely the Type-II 2HDM\cite{2HDM}, the L2HDM
\cite{l2hdm,L2HDM-phenome1}, the nMSSM and the NMSSM. In our
study, we examine carefully the constraints on the light $a$ from
many latest experimental results.

This work is organized as follows. In Sec. II we briefly describe
the four new physics models. In Sec. III we present the calculations
of the rare $Z$-decays. In Sec. IV we list the constraints on
the four new physics models. In Sec. V we show the numerical results for
the branching ratios of the rare $Z$-decays in various models.
Finally, the conclusion is given in Sec. VI.

\vspace{-0.5cm}

\section{The new physics models}
As the most economical way, the SM utilizes one Higgs doublet to
break the electroweak symmetry. As a result, the SM predicts only
one physical Higgs boson with its properties totally determined by
two free parameters. In new physics models, the Higgs sector is
usually extended by adding Higgs doublets and/or singlets, and
consequently, more physical Higgs bosons are predicted along with
more free parameters involved in.

The general 2HDM contains two $SU(2)_L$ doublet Higgs fields
$\phi_1$ and $\phi_2$, and with the assumption of CP-conserving, its
scalar potential can be  parameterized as\cite{2HDM}:
\begin{eqnarray}
V&=&m_1^2\phi_1^\dagger\phi_1 + m_2^2\phi_2^\dagger\phi_2
-(m_3^2\phi_1^\dagger\phi_2 + H.c.)
+\frac{\lambda_1}{2}(\phi_1^\dagger\phi_1)^2 + \frac{\lambda_2}{2}(\phi_2^\dagger\phi_2)^2\nonumber\\
&&+\lambda_3(\phi_1^\dagger\phi_1)(\phi_2^\dagger\phi_2)
+\lambda_4(\phi_1^\dagger\phi_2)(\phi_2^\dagger\phi_1)
+\frac{\lambda_5}{2}[(\phi_1^\dagger\phi_2)^2+H.c.],
\end{eqnarray}
where $\lambda_i$ ($i=1,\cdots,5$) are free dimensionless
parameters, and $m_i$ ($i=1,2,3$) are the parameters with mass
dimension. After the electroweak symmetry breaking, the spectrum of
this Higgs sector includes three massless Goldstone modes, which
become the longitudinal modes of $W^\pm$ and $Z$ bosons, and five
massive physical states: two CP-even Higgs bosons $h_1$ and $h_2$,
one neutral CP-odd Higgs particle $a$ and a pair of charged Higgs
bosons $H^{\pm}$. Noting the constraint $v_1^2+v_2^2=(246 ~{\rm
GeV})^2$ with $v_1$ and $v_2$ denoting the vacuum expectation values
(vev) of $\phi_1$ and $\phi_2$ respectively, we choose
\begin{eqnarray}
m_{h_1},~~~ m_{h_2}, ~~~m_a, ~~~m_{H^\pm}, ~~~\tan\beta, ~~~\sin\alpha, ~~~\lambda_5
\end{eqnarray}
as the input parameters with $\tan\beta=v_2/v_1$, and $\alpha$ being
the mixing angle that diagonalizes the mass matrix of the CP-even
Higgs fields.

The difference between the Type-II 2HDM and the L2HDM comes from the
Yukawa coupling of the Higgs bosons to quark/lepton. In the Type-II
2HDM, one Higgs doublet $\phi_2$ generates the masses of up-type
quarks and the other doublet $\phi_1$ generates the masses of
down-type quarks and charged leptons; while in the L2HDM one Higgs
doublet $\phi_1$ couples only to leptons and the other doublet
$\phi_2$ couples only to quarks. So the Yukawa interactions of $a$
to fermions in these two models are given by \cite{2HDM,l2hdm-cao}
\begin{eqnarray}
{\cal L}_{\rm Yukawa}^{\rm Type-II}&=&
 \frac{igm_{u_i}}{2m_W}\cot\beta \bar{u}_i\gamma^5u_ia
+\frac{igm_{d_i}}{2m_W}\tan\beta \bar{d}_i\gamma^5d_ia + \frac{igm_{e_i}}{2m_W}\tan\beta \bar{e}_i\gamma^5e_ia,\\
{\cal L}_{\rm Yukawa}^{\rm L2HDM}&=&\frac{igm_{u_i}}{2m_W}\cot\beta
\bar{u}_i\gamma^5u_ia -\frac{igm_{d_i}}{2m_W}\cot\beta
\bar{d}_i\gamma^5d_ia +\frac{igm_{e_i}}{2m_W}\tan\beta
\bar{e}_i\gamma^5e_ia,  \label{Yukawa}
\end{eqnarray}
with $i$ denoting generation index. Obviously, in the Type-II 2HDM
the $\bar{b} b a$ coupling and  the $\bar{\tau} \tau a$ coupling can
be simultaneously enhanced by $\tan \beta$, while in the L2HDM only
the $\bar{\tau} \tau a$ coupling is enhanced by $\tan\beta$.

The structures of the nMSSM and the NMSSM are described by their
superpotentials and corresponding soft-breaking terms, which are
given by \cite{Barger}
\begin{eqnarray}
 W_{\rm nMSSM}&=&W_{\rm MSSM} + \lambda\hat{H_u} \cdot \hat{H_d} \hat{S}
 +\xi_FM_n^2\hat S ,\\
 W_{\rm NMSSM}&=&W_{\rm MSSM} + \lambda\hat{H_u} \cdot \hat{H_d} \hat{S}
 + \frac{1}{3}\kappa \hat{S^3},  \\
V_{\rm soft}^{\rm nMSSM}&=& \tilde m_u^2|H_u|^2 + \tilde
m_d^2|H_d|^2 + \tilde m_S^2|S|^2
+(A_\lambda \lambda SH_u\cdot H_d +\xi_S M_n^3 S + h.c.),\\
V_{\rm soft}^{\rm NMSSM}&=&\tilde m_u^2|H_u|^2 + \tilde m_d^2|H_d|^2
+ \tilde m_S^2|S|^2 +(A_\lambda \lambda SH_u\cdot H_d
+\frac{A_\kappa}{3}\kappa S^3 + h.c.),
\end{eqnarray}
where $W_{\rm MSSM}$ is the superpotential of the MSSM without the $\mu$
term, $\hat{H}_{u,d}$ and $\hat S$ are Higgs doublet and singlet
superfields  with $H_{u,d}$ and $S$ being their scalar component
respectively, $\tilde m_u$, $\tilde m_d$, $\tilde m_S$, $A_\lambda$,
$A_\kappa$ and $\xi_S M_n^3 $ are soft breaking parameters, and
$\lambda$ and $\kappa$ are coefficients of the Higgs self
interactions.

With the superpotentials and the soft-breaking terms, one can get
the Higgs potentials of the nMSSM and the NMSSM respectively. Like
the 2HDM, the Higgs bosons with same CP property will mix and the
mass eigenstates are obtained by diagonalizing the corresponding
mass matrices:
\begin{eqnarray}
\left( \begin{array}{c} h_1 \\ h_2 \\ h_3 \end{array} \right) = U^H
\left( \begin{array}{c} \phi_u \\ \phi_d\\ \sigma\end{array}
\right),~ \left(\begin{array}{c} a\\ A\\ G^0 \end{array} \right) =
U^A \left(\begin{array}{c} \varphi_u \\ \varphi_d \\ \xi \end{array}
\right),~ \left(\begin{array}{c} H^+ \\G^+ \end{array}  \right) =U
\left(\begin{array}{c}H_u^+\\ H_d^+ \end{array} \right),
\end{eqnarray}
where the fields on the right hands of the equations are component
fields of $H_u$, $H_d$ and $S$ defined by
\begin{eqnarray}
H_d & =& \left ( \begin{array}{c}
             \frac{v_d + \phi_d + i \varphi_d}{\sqrt{2}}\\
             H_d^- \end{array} \right),~~
H_u = \left ( \begin{array}{c} H_u^+ \\
       \frac{v_u + \phi_u + i \varphi_u}{\sqrt{2}}
        \end{array} \right),~~
S  = \frac{1}{\sqrt{2}} \left( s + \sigma + i \xi \right),
\end{eqnarray}
$h_1,h_2,h_3$ and $a,A$ are respectively the CP-even and CP-odd
neutral Higgs bosons,  $G^0$ and $G^+$ are Goldstone bosons eaten by
$Z$ and $W^+$, and $H^+$ is the charged Higgs boson. So both the
nMSSM and NMSSM predict three CP-even Higgs bosons, two CP-odd Higgs
bosons and one pair of charged Higgs bosons. In general, the lighter
CP-odd Higgs $a$ in these model is the mixture of the singlet field
$\xi$ and the doublet field combination, $\cos \beta \varphi_u+ \sin
\beta \varphi_d$,  i.e.
\begin{eqnarray}
a = \cos \theta_A \xi + \sin \theta_A ( \cos \beta \varphi_u+ \sin
\beta \varphi_d ),   \label{mixing}
\end{eqnarray}
and its couplings to down-type quarks are then proportional to
$\frac{g m_q}{m_W} \tan \beta \sin \theta_A$. So for singlet
dominated $a$, $\sin \theta_A $ is small and the couplings are
suppressed. As a comparison, the interactions of $a$ with the
squarks are given by\cite{NMSSM}
\begin{eqnarray}
{\cal L}_{a \tilde{q}^\ast \tilde{q}} &=&  \frac{- i g m_u }{2 m_W}
\left ( \lambda v \cot \beta \cos \theta_A + A_u \cot \beta \sin
\theta_A - \mu \sin \theta_A \right ) ( \tilde{u}_R^\ast \tilde{u}_L
- \tilde{u}_L^\ast \tilde{u}_R ) a
\nonumber \\
&& - \frac{i g m_d}{2 m_W} \left ( \lambda v \tan \beta \cos
\theta_A + A_d \tan \beta \sin \theta_A - \mu \sin \theta_A \right )
( \tilde{d}_R^\ast \tilde{d}_L - \tilde{d}_L^\ast \tilde{d}_R ) a,
\label{asqsq}
\end{eqnarray}
i.e. the interaction does not vanish when $\sin \theta_A$ approaches
zero. Just like the 2HDM where we use the vevs of the Higgs fields
as fundamental parameters, we choose $\lambda$, $\kappa$, $\tan
\beta$, $\mu_{eff}= \lambda \langle S \rangle$, $A_\kappa$ and
$m_A^2 = \frac{2 \mu}{\sin 2 \beta} ( A_\lambda + \frac{\kappa
\mu}{\lambda} )$ as input parameters for the NMSSM\cite{Barger} and
$\lambda$, $\tan \beta$, $\mu_{eff}= \lambda \langle S \rangle$,
$A_\lambda$, $\tilde{m}_S$ and $m_A^2 = \frac{2}{\sin 2 \beta} ( \mu
A_\lambda + \lambda \xi_F M_n^2 )$ as input parameters for the
nMSSM\cite{nMSSM-cao}.

About the nMSSM and the NMSSM, three points should be noted. The
first is for the two models, there is no explicit $\mu -$term, and
the effective $\mu$ parameter ($\mu_{\rm eff}$) is generated when
the scalar component of $\hat S$ develops a vev. The second is,
the nMSSM is actually same as the NMSSM with
$\kappa =0$\cite{nMSSM-cao},
because the tadpole terms $\xi_FM_n^2\hat S$ and its soft breaking
term $\xi_S M_n^3 S$ in the nMSSM do not induce any interactions,
except for the tree-level Higgs boson masses and the minimization
conditions. And the last is despite of the similarities, the
nMSSM has its own peculiarity, which comes from its neutralino
sector. In the basis $(-i\lambda', - i \lambda^3, \psi_{H_u}^0,
\psi_{H_d}^0, \psi_S )$, its neutralino mass matrix is given by
\cite{nMSSM}
\begin{eqnarray}
\left( \begin{array}{ccccc}
M_1          & 0             & m_Zs_W s_b    & - m_Z s_W c_b  & 0 \\
0            & M_2           & -m_Z c_W s_b  & m_Z c_W c_b    & 0 \\
m_Zs_W s_b   & -m_Z c_W s_b  & 0             & -\mu           & -\lambda v c_b \\
-m_Z s_W c_b & m_Z c_W c_b  &  -\mu         & 0              & - \lambda v s_b \\
0            & 0             &-\lambda v c_b &- \lambda v s_b & 0
\end{array} \right)
\end{eqnarray}
where $M_1$ and $M_2$ are $U(1)$ and $SU(2)$ gaugino masses
respectively, $s_W=\sin \theta_W$, $c_W=\cos\theta_W$,
$s_b=\sin\beta$ and $c_b=\cos\beta$. After diagonalizing this matrix
one can get the mass eigenstate of the lightest neutralino
$\tilde{\chi}_1^0$ with mass taking the following form
\cite{lsp-mass}
\begin{eqnarray}
m_{\tilde{\chi}^0_1} \simeq \frac{2\mu \lambda^2 (v_u^2+
v_d^2)}{2\mu^2+\lambda^2  (v_u^2+ v_d^2)}
             \frac{\tan \beta}{\tan^2 \beta+1}
             \label{mass-exp}
\end{eqnarray}
This expression implies that $\tilde{\chi}_1^0$ must be lighter than
about $60$ GeV for $\mu > 100 {\rm GeV}$ (from lower bound on chargnio
mass) and $\lambda < 0.7$ (perturbativity bound). Like the other
supersymmetric models, $\tilde{\chi}_1^0$ as the lightest sparticle
acts as the dark matter in the universe, but due to its
singlino-dominated nature, it is difficult to annihilate
sufficiently to get the correct density in the current universe. So
the relic density of $\tilde{\chi}_1^0$ plays a crucial way in
selecting the model parameters. For example, as shown in
\cite{nMSSM-cao}, for $\tilde{\chi}_1^0 > 37 {\rm GeV}$, there is no
way to get the correct relic density, and for the other cases,
$\tilde{\chi}_1^0$ mainly annihilates by exchanging $Z$ boson for $
30 {\rm GeV} < m_{\tilde{\chi}^0_1} < 37 {\rm GeV}$ , or by
exchanging a light CP-odd Higgs boson $a$ with mass satisfying the
relation $m_a \simeq 2 m_{\tilde{\chi}^0_1}$ for
$m_{\tilde{\chi}^0_1} < 25 {\rm GeV}$. For the annihilation,
$\tan \beta$ and $\mu$ are required to be less than 10 and $500{\rm
GeV}$ respectively because through Eq.(\ref{mass-exp}) a large $\tan
\beta$ or $\mu$ will suppress $m_{\tilde{\chi}^0_1}$ to make the
annihilation more difficult. The properties of the lightest CP-odd
Higgs boson $a$, such as its mass and couplings, are also limited
tightly since $a$ plays an important role in $\tilde{\chi}_1^0$
annihilation. The phenomenology of the nMSSM is also rather special,
and this was discussed in detail in \cite{nMSSM-cao}.

\section{CALCULATIONS}

\begin{figure}[htb]
\epsfig{file=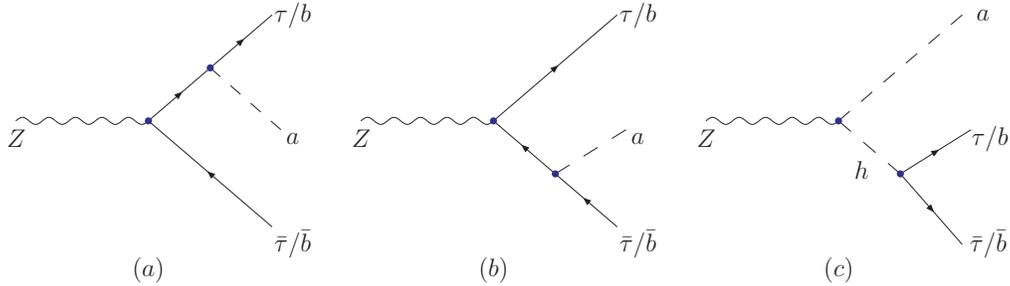,width=13.7cm} \vspace*{-0.5cm} \caption{Feynman
diagrams contributing to the decay $Z\to \bar{f} f a$ ($f=b,\tau$)
in new physics models. $h$ denotes all possible intermediate CP-even
Higgs bosons in the corresponding model.} \label{fig1}
\end{figure}

\begin{figure}[htb]
\epsfig{file=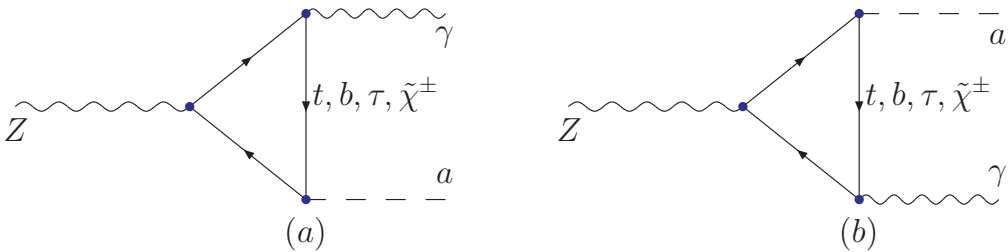,width=13.7cm} \vspace*{-0.5cm} \caption{Feynman
diagrams contributing to the decay $Z\to a\gamma$ at one-loop level
in new physics models. Note the chargino loop diagrams only exist in
the nMSSM and NMSSM.} \label{fig2}
\end{figure}

\begin{figure}[t]
\epsfig{file=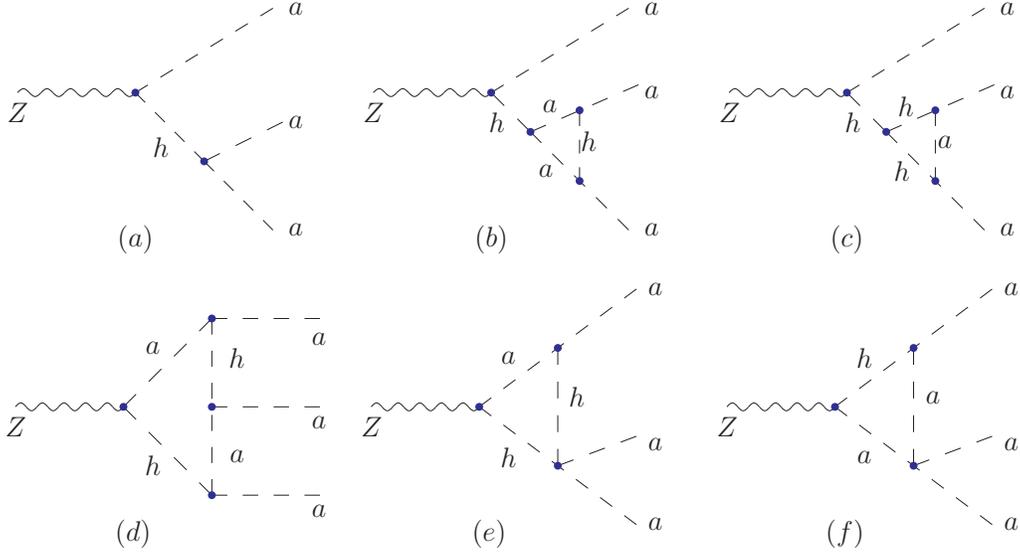,width=13.7cm} \vspace*{-0.5cm} \caption{Feynman
diagrams contributing to the decay $Z\to aaa$ in new physics models.
Only the correction from the Higgs-mediated loops are considered
since the other corrections can be safely neglected (see the
arguments in the text).} \label{fig3}
\end{figure}

In the Type-II 2HDM, L2HDM, nMSSM and NMSSM, the rare $Z$-decays $Z
\to \bar{f} f a$ ($f=b, \tau$), $Z\to a\gamma$ and $Z\to aaa$ may
proceed by the Feynman diagrams shown in Fig.\ref{fig1},
Fig.\ref{fig2} and Fig.\ref{fig3} respectively. For these diagrams,
the intermediate state $h$ represents all possible CP-even Higgs
bosons in the corresponding model, i.e. $h_1$ and $h_2$ in Type-II
2HDM and L2HDM and $h_1$, $h_2$ and $h_3$ in nMSSM and NMSSM. In
order to  take into account the possible resonance effects of $h$ in
Fig.\ref{fig1}(c) for $Z \to \bar{f} f a$ and Fig.\ref{fig3} (a) for
$Z \to a a a$, we have calculated all the decay modes of $h$ and
properly included the width effect in its propagator. As to the
decay $Z \to a \gamma$, two points should be noted. One is, unlike
the decays $Z \to \bar{f} f a $ and $Z \to a a a$, this process
proceeds only through loops mediated by quarks/leptons in the
Type-II 2HDM and L2HDM, and additionally by sparticles in the nMSSM
and NMSSM. So in most cases its rate should be much smaller than the
other two. The other is due to CP-invariance, loops mediated by
squarks/sleptons give no contribution to the
decay\cite{agamma-MSSM}. In actual calculation, this is reflected by
the fact that the coupling coefficient of $\tilde{q}_R^\ast
\tilde{q}_L a $  differs from that of  $\tilde{q}_L^\ast \tilde{q}_R
a $ by a minus sign (see Eq.(\ref{asqsq})),  and as a result, the
squark-mediated contributions to $Z \to a \gamma$ are completely
canceled out.

With regard to the rare decay $Z \to a a a$, we have more
explanations. In the lowest order, this decay proceeds by the
diagram shown in Fig.\ref{fig3} (a), and hence one may think that,
as a rough estimate, it is enough to only consider the contributions
from Fig.\ref{fig3}(a). However, we note that in some cases of the
Type-II 2HDM and L2HDM, due to the cancelation of the contributions
from different $h$ in Fig.\ref{fig3} (a) and also due to the
potentially largeness of $h a a$ couplings (i.e. larger than the
electroweak scale $v$), the radiative correction from the
Higgs-mediated loops may dominate over the tree level contribution
even when the tree level prediction of the rate, $Br_{tree} (Z \to a
a a)$, exceeds $10^{-6}$. On the other hand, we find the
contribution from quark/lepton-mediated loops can be safely
neglected if $Br_{tree} (Z \to a a a)
> 10^{-8}$ in the Type-II 2HDM and the L2HDM. In the nMSSM and the NMSSM,
besides the corrections from the Higgs- and quark/lepton-mediated
loops, loops involving sparticles such as squarks, charginos and
neutralinos can also contribute to the decay. We numerically checked
that the contributions from squarks and charginos can be safely
neglected if $Br_{tree} (Z \to a a a)
> 10^{-8}$. We also calculated part of potentially large neutralino
correction (note that there are totally about $5^4$ diagrams for
such correction!) and found they can be neglected too. Since
considering all the radiative corrections will make our numerical
calculation rather slow, we only include the most important
correction, namely that from Higgs-mediated loops, in presenting our
results for the four models.

One can intuitively understand the relative smallness of the
sparticle contribution to $Z \to a a a$ as follows. First consider
the squark contribution which is induced by the $Z\tilde{q}^\ast
\tilde{q}$ interaction ($\tilde{q}$ denotes the squark in chirality
state) and the $\tilde{q}^\ast \tilde{q} a$ interaction through box
diagrams. Because the $Z\tilde{q}^\ast \tilde{q}$ interaction
conserves the chirality of the squarks while the $\tilde{q}^\ast
\tilde{q} a$ interaction violates the chirality, to get non-zero
contribution to $Z \to a a a$ from the squark loops, at least four
chiral flippings are needed, with three of them provided by
$\tilde{q}^\ast \tilde{q} a$ interaction and the rest provided by
the left-right squark mixing. This means that, if one calculates the
amplitude in the chirality basis with the mass insertion method, the
amplitude is suppressed by the mixing factor $\frac{m_q
X_q}{m_{\tilde{q}}^2}$ with $m_q X_q$ being the off diagonal element
in squark mass matrix. Next consider the chargino/neutralino
contributions. Since for a light $a$, its doublet component,
parameterized by $\sin \theta_A $ in Eq.(\ref{mixing}), is usually
small, the couplings of $a$ with the sparticles will never be
tremendously large\cite{NMSSM}. So the chargino/neutralino
contributions are not important too. In our calculation of the
decays, we work in the mass eigenstates of sparticles instead of in
the chirality basis.

\vspace{-0.2cm}

\section{Constraints on the new physics models}

For the Type-II 2HDM and the L2HDM, we consider the following
constraints \cite{l2hdm-cao}:
\begin{itemize}
\item[(1)] Theoretical constraints on $\lambda _i$ from perturbativity, unitarity
and requirements that the scalar potential is finit at large field
values and contains no flat directions \cite{2HDM,Unitary}, which
imply that
\begin{eqnarray}
&&\lambda _i < 4 \pi ~(i=1,5),~~ \lambda_{1,2}>0,~~
\lambda_3 > -\sqrt{\lambda_1\lambda_2},~~
\lambda_3 +\lambda_4 -|\lambda_5| > -\sqrt{\lambda_1\lambda_2},\nonumber\\
&&3(\lambda_1 +\lambda_2) \pm
 \sqrt{9(\lambda_1 -\lambda_2)^2 +4(2\lambda_3 +\lambda_4)^2}< 16\pi,\nonumber\\
&&\lambda_1 +\lambda_2 \pm
 \sqrt{(\lambda_1 -\lambda_2)^2 +4|\lambda_4|^2}< 16\pi,\nonumber\\
&&\lambda_1 +\lambda_2 \pm
 \sqrt{(\lambda_1 -\lambda_2)^2 +4|\lambda_5|^2}< 16\pi,\nonumber\\
&&\lambda_3 +2\lambda_4 \pm 3|\lambda_5|< 8\pi,~~
\lambda_3 \pm \lambda_4 < 8\pi,~~ \lambda_3 \pm |\lambda_5| < 8\pi.
\end{eqnarray}

\item[(2)] The constraints from the LEP search for neutral Higgs
bosons. We compute the signals from the Higgs-strahlung production
$e^+ e^- \to Z h_i $ ($i= 1, 2$) with $h_i \to 2 b, 2 \tau, 4 b, 4
\tau, 2 b 2 \tau $ \cite{LHWGSM,Opal3,ALEPH1} and from the
associated production $e^+ e^- \to h_i a$ with $h_i a \to 4 b, 4
\tau, 2 b 2 \tau, 6 b, 6 \tau$ \cite{DELPHI}, and compare them with
the corresponding LEP data which have been inputted into our code.
We also consider the constraints from $e^+ e^- \to Z h_i $ by
looking for a peak of $M_{h_i}$ recoil mass distribution of
$Z$-boson \cite{ALEPH} and the constraint of $\Gamma ( Z \to h_i a)
< 5.8$ MeV when $m_a + m_{h_i} < m_Z $ \cite{Monig}.

These constraints limit the quantities such as
$C_{eff}^{2b}=[g_{ZZh_i}^2/g_{ZZh_{SM}}^2] \times Br (h_i \to
\bar{b} b )$  on the $C_{eff}^{2b}-m_{h_i}$ plane with the the
subscript $g_{ZZh_i}$ denoting the coupling coefficient of the
$ZZh_i$ interaction. They also impose a model-dependent lower bound
on $m_{h_i}$, e.g., $70 {\rm GeV}$ for the Type-II 2HDM (from
our scan results), $50 {\rm
GeV}$ for the L2HDM\cite{l2hdm-cao}, and $30 {\rm GeV}$ for the
nMSSM \cite{nMSSM-cao}. These bounds are significantly lower than
that of the SM, i.e. $114 {\rm GeV}$, partially because in new
physics models, unconventional decay modes of $h_i$ such as $h_i \to
a a $ are open up. As to the nMSSM, another specific reason for
allowing a significantly lighter CP-even Higgs boson is that the
boson may be singlet-dominated in this model.

With regard to the lightest CP-odd Higgs boson $a$, we checked that
there is no lower bound on its mass so long as the $Z h_i a$
interaction is weak or $h_i$ is sufficiently heavy.

\item[(3)] The constraints from the LEP search for a light Higgs boson
via the Yukawa process $e^+ e^- \to \bar{f} f S$ with $f=b, \tau$
and $S$ denoting a scalar \cite{ffS}. These constraints can limit
the  $\bar{f} f S$ coupling versus $m_S$ in new physics models.

\item[(4)] The constraints from the CLEO-III limit on
$Br ( \Upsilon (1S) \to a \gamma \to  \tau^+ \tau^- \gamma)$ and the
latest BaBar limits on $Br(\Upsilon(3S) \to a \gamma \to \tau^+
\tau^- \gamma, \mu^+ \mu^- \gamma)$. These constraints will put very
tight constraints on the $a\bar{b}b$ coupling for $m_a < 9 {\rm
GeV}$. In our analysis, we use the results of Fig.8 in the second paper
 of \cite{Gunion} to excluded the unfavored points.

\item[(5)] The constraints from $Z \tau^+ \tau^-$ couplings.
Since the Higgs sector can give sizable higher order corrections to
$Z \tau^+ \tau^-$ couplings, we calculate them to one loop level and
require the corrected $Z \tau^+ \tau^-$ couplings to lie within the
$2 \sigma $ range of their fitted value. The SM predictions for the
couplings at $Z$-pole are given by $g_V^{SM}=-0.03712 $ and
$g_A^{SM} = -0.50127 $ \cite{LEP-Report}, and the fitted values are
given by $ -0.0366 \pm 0.00245 $ and $ -0.50204 \pm 0.00064 $,
respectively\cite{LEP-Report}. We adopt the formula in \cite{Cao}
to the 2HDM in our calculation.

\item[(6)]  The constraints from $\tau$ leptonic decay.
We require the new physics correction to the branching ratio $Br(
\tau \to e \bar{\nu}_e \nu_\tau)$ to be in the range of $-0.80\%
\sim 1.21\%$ \cite{tau-decay}. We use the formula in
\cite{tau-decay} in our calculation.

About the constraints (5) and (6), two points should be noted. One
is all Higgs bosons are involved in the constraints by entering the
self energy of $\tau$ lepton, the $Z \bar{\tau} \tau$ vertex
correction or the $W \bar{\tau} \nu_\tau$ vertex correction, and
also the box diagrams for $\tau \to e \bar{\nu}_e
\nu_\tau$\cite{Cao,tau-decay}. Since the Yukawa couplings of the
Higgs bosons to $\tau$ lepton get enhanced by $\tan \beta$ and so do
the corrections, $\tan \beta$ must be upper bounded for given
spectrum of the Higgs sector. Generally speaking, the lighter $a$
is, the more tightly $\tan \beta$ is
limited\cite{tau-decay,l2hdm-cao}. The other point is in the Type-II
2HDM, $R_b$, B-physics observables as well as $\Upsilon$ decays
discussed above can constraint the model in a tighter way than the
constraints (5) and (6) since the Yukawa couplings of $\tau$ lepton
and $b$ quark are simultaneously enhanced by $\tan \beta$. But for
the L2HDM, because only the Yukawa couplings of $\tau$ lepton get
enhanced (see Eq.\ref{Yukawa}), the constraints (5) and (6) are more
 important in limiting $\tan \beta$.

\item[(7)] Indirect constraints from the precision electroweak
observables such as $\rho_{\ell}$, $\sin^2 \theta_{eff}^{\ell}$ and
$M_W$, or their combinations $\epsilon_i (i=1,2,3)$
\cite{Altarelli}. We require $\epsilon_i$ to be compatible with the
LEP/SLD data at $95\%$ confidence level\cite{LEP-Report}. We also
require new physics prediction of $R_b= \Gamma (Z \to \bar{b} b) /
\Gamma ( Z \to {\rm hadrons} )$ is within the $2 \sigma$ range of
its experimental value. The latest results for $R_b $ are $R_b^{exp}
= 0.21629 \pm 0.00066 $ (measured value) and $R_b^{SM} = 0.21578 $
(SM prediction) for $m_t = 173$ GeV \cite{PDG}.  In our code, we
adopt the formula  for these observables presented in \cite{Cao} to
the Type-II 2HDM and the L2HDM respectively.

In calculating $\rho_{\ell}$, $\sin^2 \theta_{eff}^{\ell}$ and
$M_W$, we note that these observables get dominant contributions
from the self energies of the gauge bosons $Z$, $W$ and $\gamma$.
Since there is no $Z a a$ coupling or $\gamma a a$ coupling, $a$
must be associated with the other Higgs bosons to contribute to the
self energies. So by the UV convergence of these quantities, one can
infer that, for the case of a light $a$ and $m_{h_i}, m_{H^\pm} \gg
m_Z$, these quantities depend on the spectrum of the Higgs sector in
a way like $\ln\frac{m_{h_i}^2}{m_{H^\pm}^2}$ at leading order,
which implies that a light $a$ can still survive the constraints
from the precision electroweak observables given the splitting
between $m_{h_i}$ and $m_{H^\pm}$ is moderate\cite{l2hdm-cao}.

\item[(8)] The constraints from B physics observables
such as the branching ratios for $B \to X_s \gamma$, $B_s \to \mu^+
\mu^-$ and $B^+ \to \tau^+ \nu_\tau$, and the mass differences
$\Delta M_d$ and $\Delta M_s$.  We require their theoretical
predications to agree with the corresponding experimental values at
$2 \sigma$ level.

In the Type-II 2HDM and the L2HDM, only the charged Higgs boson
contributes to these observables by loops, so one can expect that
$m_{H^\pm}$ versus $\tan \beta$ is to be limited. Combined analysis
of the limits in the Type-II 2HDM has been done by the CKMfitter
Group, and the lower bound of $m_{H^{\pm}}$ as a function of $\tan
\beta$ was given in Fig.11 of \cite{2HDM-charge}. This analysis
indicates that $m_{H^\pm}$ must be heavier than $ 316 {\rm GeV}$ at
$95\%$ C.L. regardless the value of $\tan \beta$. In this work, we
use the results of Fig.11 in \cite{2HDM-charge} to exclude the
unfavored points. As for the L2HDM, B physics actually can not put
any constraints\cite{Su} because in this model the couplings of the
charged Higgs boson to quarks are proportional to $\cot \beta$ and
in the case of large $\tan \beta$ which we are interested in, they
are suppressed. In our analysis of the L2HDM, we impose the LEP
bound on $m_{H^\pm}$, i.e. $m_{H^\pm} > 92 {\rm
GeV}$\cite{OPALchargedHiggs}.

\item[(9)]The constraints from the muon anomalous magnetic moment $a_\mu$. Now
both the theoretical prediction and the experimental measured value
of $a_\mu$ have reached a remarkable precision, but a significant
deviation still exists: $a_\mu^{exp} - a_\mu^{SM} = (25.5 \pm 8.0 )
\times 10^{-10}$ \cite{Davier}. In the 2HDM,  $a_\mu$ gets
additional contributions from the one-loop diagrams induced by the
Higgs bosons and also from the two-loop Barr-Zee diagrams mediated
by $a$ and $h_i$\cite{Barr-Zee}. If the Higgs bosons are much
heavier than $\mu$ lepton mass, the contributions from the Barr-Zee
diagrams are more important, and to efficiently alleviate the
discrepancy of $a_\mu$, one needs a light $a$ along with its
enhanced couplings to $\mu$ lepton and also to heavy fermions such
as bottom quark and $\tau$ lepton to push up the effects of the
Barr-Zee diagram\cite{Barr-Zee}. The CP-even Higgs bosons are
usually preferred to be heavy since their contributions to $a_\mu$
are negative.

In the Type-II 2HDM, because $\tan \beta$ is tightly constrained by
the process $e^+e^- \to \bar{b} b a$ at the LEP\cite{ffS} and the
$\Upsilon$ decay\cite{Gunion}, the Barr-Zee diagram contribution is
insufficient to enhance $a_\mu$ to $2 \sigma$ range around its
measured value\cite{2HDM-mu2}. So in our analysis, we require the
Type-II 2HDM to explain $a_\mu$ at $3 \sigma$ level. While for the
L2HDM, $\tan \beta$ is less constrained compared with the Type-II
2HDM, and the Barr-Zee diagram involving the $\tau$-loop is capable
to push up greatly the theoretical prediction of
$a_\mu$\cite{l2hdm-cao}. Therefore, we require the L2HDM to explain
the discrepancy at $2 \sigma$ level.

Unlike the other constraints discussed above, the $a_\mu$ constraint
will put a two-sided bound on $\tan \beta$ since on the one hand, it
needs a large $\tan \beta$ to enhance the Barr-Zee contribution, but
on the other hand, too large $\tan \beta$ will result in an
unacceptable large $a_\mu$.

\item[(10)] Since this paper concentrates on a light $a$,
the decay $h_i \to a a$ is open up with a possible large decay
width. We require the width of any Higgs boson to be smaller than
its mass to avoid a too fat Higgs boson\cite{yuan}. We checked that
for the scenario characterized by $ m_{h_2}/m_{h_1} > 3 $, the
coefficient of $h_i a a$ interaction is usually larger than the
electroweak scale $v$, and consequently a large decay width is
resulted.
\end{itemize}

For the nMSSM and NMSSM, the above constraints become more
complicated because in these models, not only more Higgs bosons are
involved in, but also sparticles enter the constraints. So it is not
easy to understand some of the constraints intuitively. Take the
process $B \to X_s \gamma$ as an example. In the supersymmetric
models, besides the charged Higgs contribution, chargino loops,
gluino loops as well as neutralino loops also contribute to the
process\cite{bsr}, and depending on the SUSY parameters, any of
these contributions may become dominated over or be canceled by
other contributions. As a result, although the charged Higgs affects
the process in the same way as that in the Type-II 2HDM, charged
Higgs as light as $130{\rm GeV}$ is still allowed even for $\tan
\beta > 50$\cite{Cao-MSSM}.

Since among the constraints, $a_\mu$ is rather peculiar in that it
needs new physics to explain the discrepancy between $a_\mu^{exp}$
and $a_\mu^{SM}$, we discuss more about its dependence on SUSY
parameters. In the nMSSM and the NMSSM, $a_\mu$ receives
contributions from Higgs loops and neutralino/chargino loops. For
the Higgs contribution, it is quite similar to that of the Type-II
2HDM except that more Higgs bosons are involved in\cite{Gunion-g2}.
For the neutralino/chargino contribution, in the light bino limit
(i.e. $M_1 \ll M_2, \mu$), it can be approximated by\cite{Martin}
\begin{eqnarray}
\delta a_\mu = 18 \tan \beta \left ( \frac{100 {\rm
GeV}}{m_{\tilde{\mu}}} \right )^3 \left ( \frac{\mu - A_t \cot
\beta}{1000 {\rm GeV}} \right ) 10^{-10}
\end{eqnarray}
for $m_{\tilde{\mu}_1} \simeq m_{\tilde{\mu}_2} = m_{\tilde{\mu}} =
2 M_1$ with $m_{\tilde{\mu}_i}$ being smuon mass. So combining the
two contributions together, one can learn that a light $a$ along
with large $\tan \beta$ and/or light smuon with moderate $\tan
\beta$ are favored to dilute the discrepancy.

Because more parameters are involved in the constraints on the
supersymmetric models, we consider following additional constraints
to further limit their parameters:
\begin{itemize}
\item[(a)] Direct bounds on sparticle masses from the LEP1, the LEP2 and the Tevatron
       experiments \cite{PDG}.
\item[(b)] The LEP1 bound on invisible Z decay
     $\Gamma(Z\to \tilde\chi_1^0 \tilde\chi_1^0) < 1.76~{\rm MeV}$;
    the LEP2 bound on neutralino production
    $\sigma(e^+e^-\to \tilde\chi_1^0 \tilde\chi_i^0) < 10^{-2}~{\rm pb}~ (i>1)$
    and $\sigma(e^+e^-\to \tilde\chi_i^0 \tilde\chi_j^0) < 10^{-1}~{\rm  pb}~ (i,j>1)$\cite{Abdallah}.
\item[(c)] Dark matter constraints from the WMAP relic density
     0.0975 $< \Omega h^2 <$ 0.1213 \cite{Dunkley}.
\end{itemize}

Note that among the above constraints, the constraint (2) on Higgs
sector and the constraint (c) on neutralino sector are very
important. This is because in the supersymmetric models, the SM-like
Higgs is upper bounded by about $100 {\rm GeV}$ at tree level and by
about $140 {\rm GeV}$ at loop level, and that the relic density
restricts the LSP annihilation cross section in a certain narrow
range.

In our analysis of the NMSSM, we  calculate the constraints (3) and
(5-7) by ourselves and utilize the code NMSSMTools \cite{NMSSMTools}
to implement the rest constraints. We also extend NMSSMTools to the
nMSSM to implement the constraints. For the extension, the most
difficult thing we faced is how to adapt the code
micrOMEGAs\cite{micrOMEGAs} to the nMSSM case. We solve this problem
by noting the following facts:
\begin{itemize}
\item As we mentioned before, the nMSSM is actually same as the
NMSSM with the trilinear singlet term setting to zero. So we can
utilize the model file of the NMSSM as the input of the micrOMEGAs
and set $\kappa = 0$.
\item Since in the nMSSM, the LSP is too light to annihilate into
Higgs pairs, there is no need to reconstruct the effective Higgs
potential to calculate precisely the annihilation channel $\chi^0_1
\chi^0_1 \to S S$ with $S$ denoting any of Higgs
bosons\cite{Belanger}.
\end{itemize}
We thank the authors of the NMSSMTools for helpful discussion on
this issue when we finish such extension\cite{nMSSM-cao}.

\section{Numerical results and discussions}

With the above constraints, we perform four independent random scans
over the parameter space of the Type-II 2HDM, the L2HDM, the nMSSM
and the NMSSM respectively. We vary the parameters in following
ranges:
\begin{eqnarray}
1\leq\tan\beta \leq80, ~-\sqrt{2}/2\leq\sin\alpha \leq\sqrt{2}/2,
~m_a\leq 30~ {\rm GeV},~ \lambda_5\leq 4\pi,\nonumber\\
5 ~{\rm GeV}\leq m_{h_1,h_2}\leq 500 ~{\rm GeV},~~~~
 316 ~{\rm GeV} \leq m_{H^+}\leq 500 ~{\rm GeV}~~~~~
\end{eqnarray}
for the Type-II 2HDM,
\begin{eqnarray}
1\leq\tan\beta \leq80, ~-\sqrt{2}/2\leq\sin\alpha \leq\sqrt{2}/2,
~m_a\leq 30~ {\rm GeV},~ \lambda_5\leq 4\pi,\nonumber\\
5 ~{\rm GeV}\leq m_{h_1,h_2}\leq 500 ~{\rm GeV},~~~~ 92 {\rm
GeV}\leq m_{H^+}\leq 500 ~{\rm GeV}~~~~~
\end{eqnarray}
for the L2HDM,
\begin{eqnarray}
0.1\leq\lambda \leq0.7,~1 \leq\tan\beta\leq 80,~100~{\rm GeV}\leq
m_A\leq 1~{\rm TeV},
  \nonumber\\
50~ {\rm GeV}\leq \mu_{\rm eff}, M_1 \leq 500~ {\rm GeV},~ -1 ~{\rm
TeV}\leq A_\lambda \leq 1 ~{\rm TeV}, ~ 0 \leq \tilde{m}_S \leq 200
{\rm GeV}
\end{eqnarray}
for the nMSSM, and
\begin{eqnarray}
0.1\leq\lambda, \kappa\leq0.7,~1 \leq\tan\beta\leq 80,~
 100~{\rm GeV}\leq m_A\leq 1~{\rm TeV},\nonumber\\
50~ {\rm GeV}\leq \mu_{\rm eff}, M_1\leq 500~ {\rm GeV},~ -100 ~{\rm
GeV}\leq A_\kappa\leq 100 ~{\rm GeV}
\end{eqnarray}
for the NMSSM.

In performing the scans, we note that for the nMSSM and the NMSSM,
some constraints also rely on the gaugino masses and the soft
breaking parameters in the squark sector and the slepton sector.
Since these parameters affect little on the properties of $a$, we
fix them to reduce the number of free parameters in our scan. For
the squark sector, we adopt the $m_h^{max}$ scenario which assumes
that the soft mass parameters for the third generation squarks are
degenerate: $M_{Q_3}=M_{U_3}=M_{D_3}=$ 800 GeV, and that the
trilinear couplings of the third generation squarks are also
degenerate, $A_t=A_b$ with  $X_t=A_t-\mu \cot\beta= -2 M_{Q_3}$. For
the slepton sector, we assume all the soft-breaking masses and
trilinear parameters to be 100 GeV. This setting is necessary for
the nMSSM since this model is difficult to explain the muon
anomalous moment at $2\sigma$ level for heavy
sleptons\cite{nMSSM-cao}. Finally, we assume the grand unification
relation $3 M_1/5\alpha_1=M_2/\alpha_2=M_3/\alpha_3$ for the gaugino
masses with $\alpha_i$ being fine structure constants of the
different gauge group.

With large number of random points in the scans, we finally get
about $3000$, $5000$, $800$ and $3000$ samples for the Type-II 2HDM,
the L2HDM, the nMSSM and the NMSSM respectively which survive the
constraints and satisfy $m_a \leq 30 {\rm GeV}$. Analyzing the
properties of the $a$ indicates that for most of the surviving
points in the nMSSM and the NMSSM, its dominant component is the
singlet field (numerically speaking, $\cos \theta_A > 0.7$) so that
its couplings to the SM fermions are
suppressed\cite{light-a-NMSSM,nMSSM-cao}. Our analysis also
indicates that the main decay products of $a$ are $\bar{\tau} \tau$
for the L2HDM\cite{l2hdm-cao}, $\bar{b} b$ (dominant) and
$\bar{\tau} \tau$ (subdominant) for the Type-II 2HDM, the nMSSM and
the NMSSM, and in some rare cases, neutralino pairs in the
nMSSM\cite{nMSSM-cao}.

\begin{figure}[t]
\epsfig{file=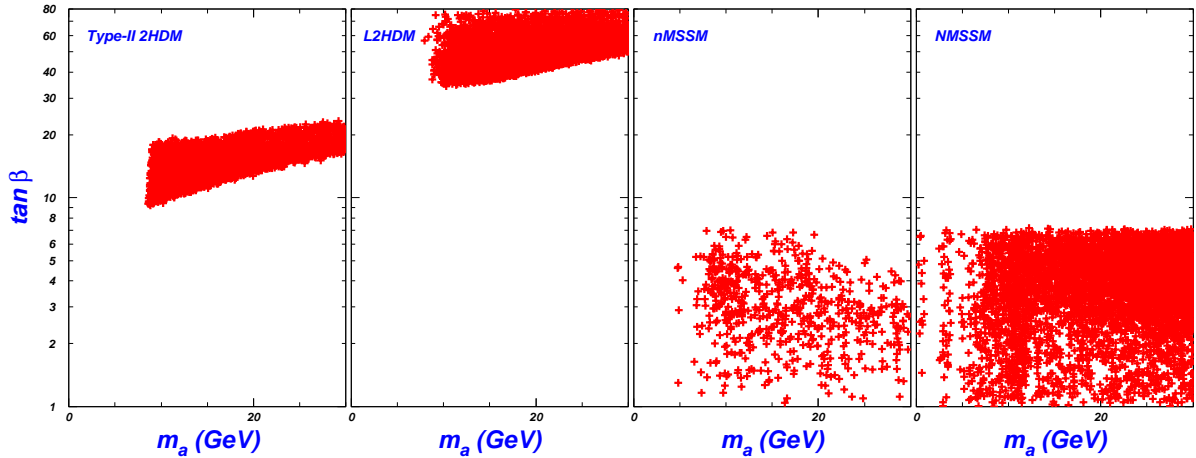,width=16cm} \vspace*{-0.5cm} \caption{The
scattering plot of the surviving samples projected on
$m_a-\tan\beta$ plane.} \label{fig4}
\end{figure}

In Fig.\ref{fig4}, we project the surviving samples on the
$m_a-\tan\beta$ plane. This figure shows that the allowed range of
$\tan \beta$ is from $8$ to $20$ in the Type-II 2HDM, and from $37$
to $80$ in the L2HDM. Just as we introduced before, the lower bounds
of $\tan \beta $ come from the fact that we require the models to
explain the muon anomalous moment, while the upper bound is due to
we have imposed the constraint from the LEP process  $e^+ e^- \to
\bar{b} b S \to 4 b$, which have limited the upper reach of the
$\bar{b} b S$ coupling for light $S$ \cite{ffS}(for the dependence
of $\bar{b} b S$ coupling on $\tan \beta$, see Sec. II). This figure
also indicates that for the nMSSM and the NMSSM, $\tan \beta$ is
upper bounded by $10$. For the nMSSM, this is because large $\tan
\beta$ can suppress the dark matter mass to make its annihilation
difficult (see \cite{nMSSM-cao} and also Sec. II), but for the
NMSSM, this is because we choose a light slepton mass so that large
$\tan \beta$ can enhance $a_\mu$ too significantly to be
experimentally unacceptable. We checked that for the slepton mass as
heavy as $300 {\rm GeV}$, $\tan \beta \geq 25 $ is still allowed for
the NMSSM.

\begin{figure}[t]
\epsfig{file=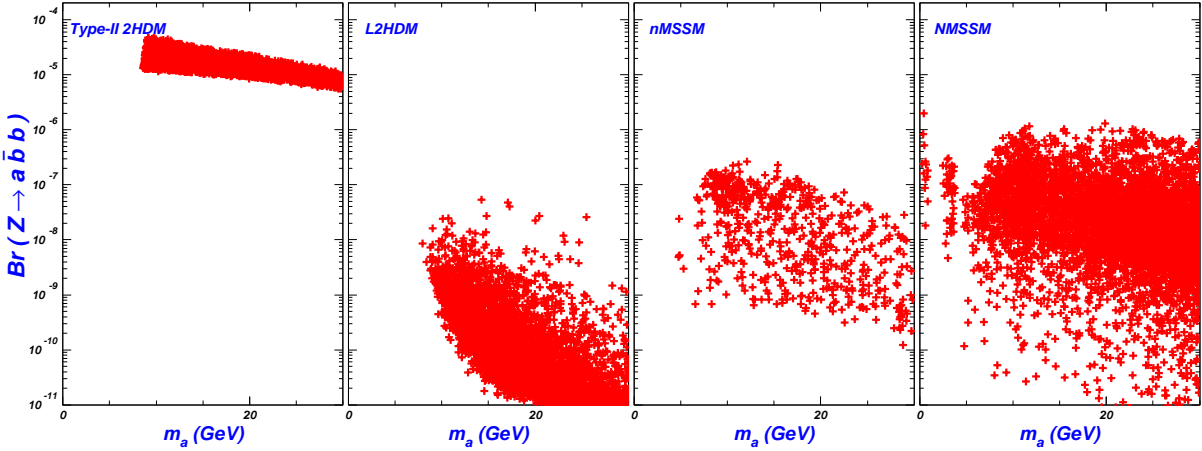,width=16cm} \vspace*{-0.5cm} \caption{Same as
Fig.\ref{fig4}, but for the branching ratio of
         $Z\to \bar{b} b a$ versus $m_a$.}
\label{fig5}
\end{figure}

\begin{figure}[t]
\epsfig{file=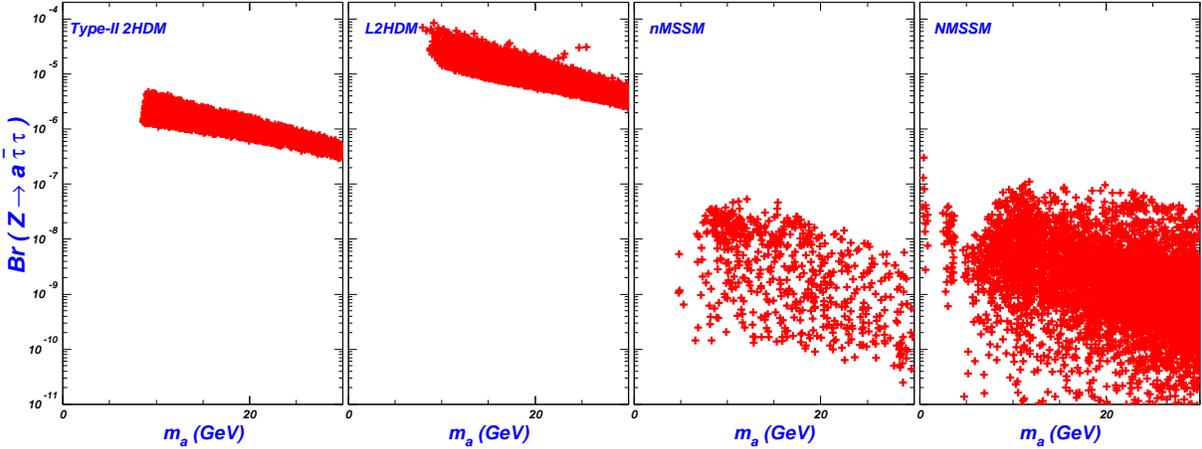,width=16cm} \vspace*{-0.5cm} \caption{Same as
Fig.\ref{fig5}, but for  $Z\to \bar{\tau} \tau a$.} \label{fig6}
\end{figure}

In Fig.\ref{fig5} and Fig.\ref{fig6}, we show the branching ratios
of $Z \to \bar{b} b a $ and $Z \to \bar{\tau} \tau a $ respectively.
Fig.\ref{fig5} indicates, among the four models, the Type-II 2HDM
predicts the largest ratio for $Z \to \bar{b} b a $ with its value
varying from $5 \times 10^{-6}$ to $ 6 \times 10^{-5}$. The
underlying reason is in the Type-II 2HDM, the $\bar{b} b a $
coupling is enhanced by $\tan \beta$ (see Fig.\ref{fig4}), while in
the other three model, the coupling is suppressed either by $\cot
\beta$ or by the singlet component of the $a$. Fig.\ref{fig6} shows
that the L2HDM predicts the largest rate for $Z \to \bar{\tau} \tau
a $ with its value reaching $10^{-4}$ in optimum case, and for the
other three models, the ratio of $Z \to \bar{\tau} \tau a $ is at
least about one order smaller than that of $Z \to \bar{b} b a$. This
feature can be easily understood from the $\bar{\tau} \tau a $
coupling introduced in Sect. II. Here we emphasize that, if the
nature prefers a light $a$, $Z \to \bar{b} b a $ and/or $Z \to
\bar{\tau} \tau a$ in the Type-II 2HDM and the L2HDM will  be
observable at the GigaZ. Then by the rates of the two decays, one
can determine whether the Type-II 2HDM or the L2HDM is the right
theory. On the other hand, if both decays are observed with small
rates or fail to be observed, the singlet extensions of the MSSM are
favored.

\begin{figure}[t]
\epsfig{file=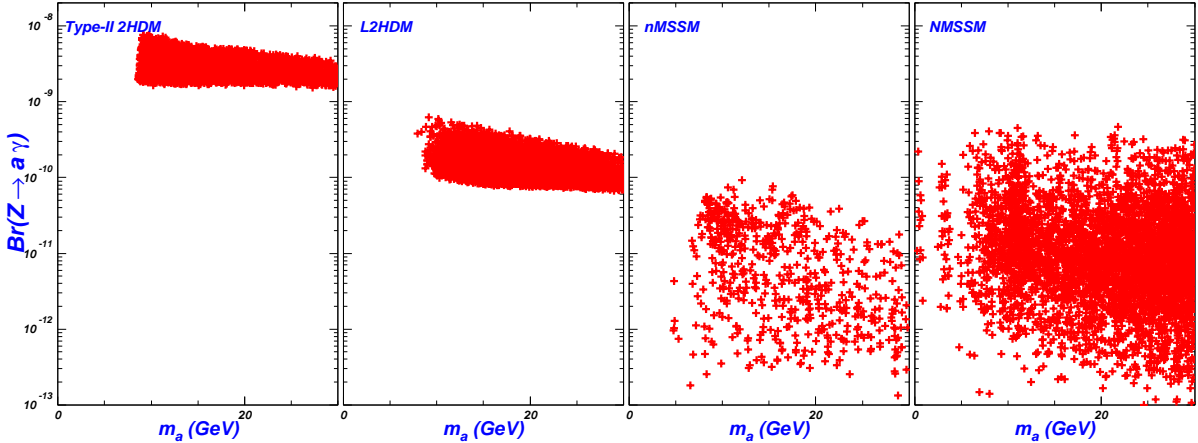,width=16cm} \vspace*{-0.5cm} \caption{Same as
Fig.\ref{fig5}, but for $Z\to a \gamma$.} \label{fig7}
\end{figure}

\begin{figure}[t]
\epsfig{file=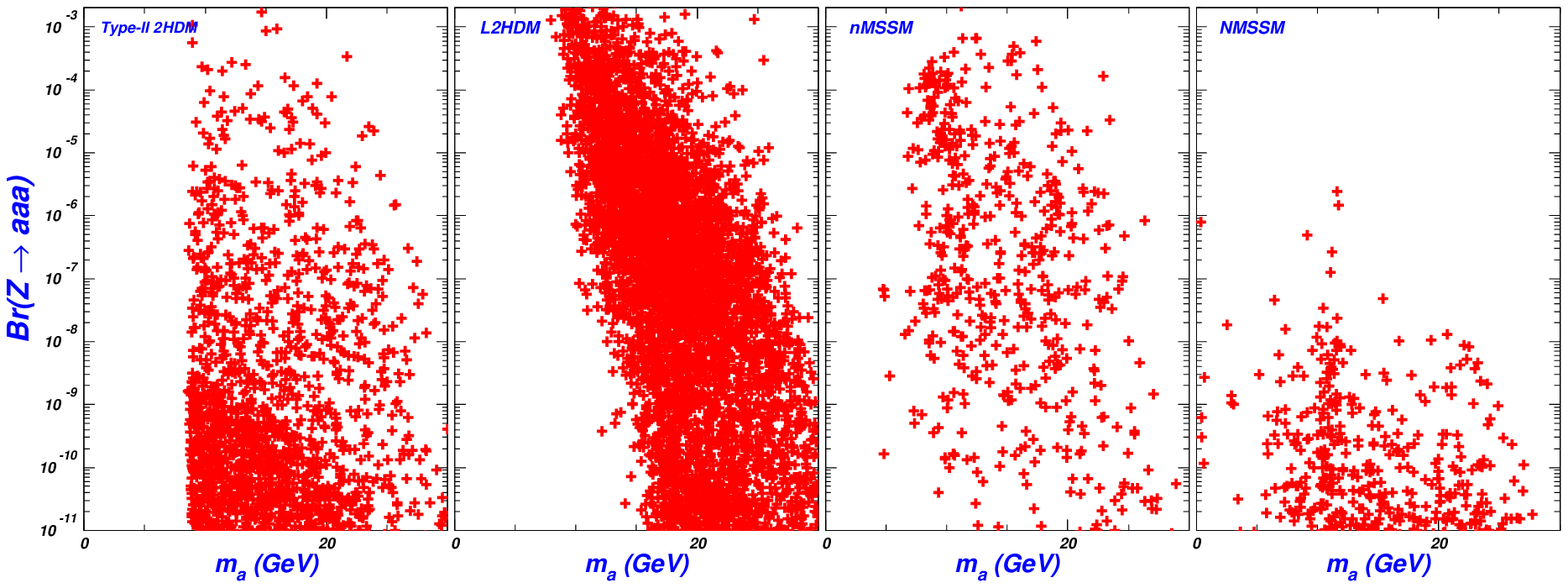,width=16cm} \vspace*{-0.5cm} \caption{Same as
Fig.\ref{fig5}, but for  $Z\to aaa $.} \label{fig8}
\end{figure}

In Fig.\ref{fig7}, we show the rate of $Z\to a\gamma$ as the
function of $m_a$. This figure indicates that the branching ratio of
$Z \to a \gamma$ can reach $9 \times 10^{-9}$, $6 \times 10^{-10}$,
$9 \times 10^{-11}$ and $4 \times 10^{-10}$ for the optimal cases of
the Type-II 2HDM, the L2HDM, the nMSSM and the NMSSM respectively,
which implies that the decay $Z \to a \gamma$ will never be
observable at the GigaZ if the studied model is chosen by nature.
The reason for the smallness is, as we pointed out before, that the
decay $Z \to a \gamma$ proceeds only at loop level.

Comparing the optimum cases of the Type-II 2HDM, the nMSSM and the
NMSSM shown in Fig.5-7, one may find that the relation $Br_{2HDM}
> Br_{NMSSM} > Br_{nMSSM}$ holds for any of the decays. This is
because the decays are all induced by the Yukawa couplings with
similar structure for the models. In the supersymmetric models, the
large singlet component of the light $a$ is to suppress the Yukawa
couplings, and the $a$ in the nMSSM has more singlet component than
that in the NMSSM.

Next we consider the decay $Z \to a a a$, which, unlike the above
decays, depends on the Higgs self interactions. In Fig.\ref{fig8} we
plot its rate as a function of $m_a$ and this figure indicates that
the $ Br( Z \to a a a )$ may be the largest among the ratios of the
exotic $Z$ decays, reaching $ 10^{-3}$ in the optimum cases of the
Type-II 2HDM, the L2HDM and the nMSSM. The underlying reason is, in
some cases, the intermediate state $h$ in Fig.\ref{fig3} (a) may be
on-shell. In fact, we find this is one of the main differences
between the nMSSM and the NMSSM, that is, in the nMSSM, $h$ in
Fig.\ref{fig3} (a) may be on-shell (corresponds to the points with
large $Br(Z \to a a a)$) while in the NMSSM, this seems impossible.
So we conclude that the decay $Z \to a a a$ may serve as an
alternative channel to test new physics models, especially it may be
used to distinguish the nMSSM from the NMSSM if the supersymmetry is
found at the LHC and the $Z \to a a a$ is observed at the GigaZ with
large rate.

Before we end our discussion, we note that in the NMSSM, the Higgs
boson $a$ may be lighter than $1 {\rm GeV}$ without conflicting with
low energy data from $\Upsilon$ decays and the other observables
(see Fig.\ref{fig4}-\ref{fig8}). In this case, $a$ is axion-like as
pointed out in \cite{light-a-NMSSM-1}. We checked that, among the
rare $Z$ decays discussed in this paper, the largest branching ratio
comes from $Z \to a \bar{b} b$ which can reach $1.9 \times 10^{-6}$.
Since in this case, the decay product of $a$ is highly collinear
muon pair, detecting the decay $Z \to a \bar{b} b$ may need some
knowledge about detectors, which is beyond our discussion.

\vspace*{-0.6cm}

\section{CONCLUSION}
In this paper, we studied the rare $Z$-decays $Z \to \bar{f} f a$
($f=b,\tau$), $Z\to a \gamma$ and $Z\to aaa$ in the Type-II 2HDM,
lepton-specific 2HDM, nMSSM and NMSSM, which predict a light CP-odd
Higgs boson $a$. In the parameter space allowed by current
experiments, the branching ratio can be as large as $10^{-4}$ for $Z
\to \bar{f} f a$, $10^{-9}$ for $Z\to a\gamma$ and $10^{-3}$ for
$Z\to aaa$, which implies that the decays $Z \to \bar{f} f a$ and $Z
\to a a a$ may be accessible at the GigaZ option. Since different
models predict different size of branching ratios, these decays can
be used to distinguish different model through the measurement of
these rare decays.

\vspace*{-0.6cm}

\section*{Acknowledgment}
This work was supported in part by HASTIT under grant No.
2009HASTIT004, by the National Natural Science Foundation of China
(NNSFC) under grant Nos. 10821504, 10725526, 10635030, 10775039,
11075045 and by the Project of Knowledge Innovation Program (PKIP)
of Chinese Academy of Sciences under grant No. KJCX2.YW.W10.


\vspace*{-0.2cm}

\begin{thebibliography}{99}

\bibitem{gigaz} J. A. Aguilar-Saavedra {\it et al.}, hep-ph/0106315.

\bibitem{Z-review} For some reviews, see, e.g.,
   M.~A.~Perez, G.~Tavares-Velasco and J.~J.~Toscano,
  Int.\ J.\ Mod.\ Phys.\  A {\bf 19}, 159 (2004);
  J. M. Yang, arXiv:1006.2594.

\bibitem{Z-decay-susy}
    J.~I.~Illana, M.~Masip, \PRD67, 035004 (2003);
    J. Cao, Z. Xiong, J. M. Yang, \EPJC32, 245 (2004).

\bibitem{Z-decay-qq}
    D. Atwood {\it et al}., \PRD66, 093005 (2002).

\bibitem{ffa} J. Kalinowski, and S. Pokorski, \PLB219, 116 (1989);
A. Djouadi, P. M. Zerwas and J. Zunft, \PLB259, 175 (1991); A.
Djouadi, J. Kalinowski, and P. M. Zerwas, Z. Phys. C {\bf 54}, 255
(1992).

\bibitem{agamma-2HDM} M. Krawczyk, {\it et al.}, \EPJC19, 463 (2001); \EPJC8, 495 (1999).

\bibitem{agamma-MSSM} J. F. Gunion, G. Gamberini and S. F. Novaes, \PRD38, 3481 (1988);
Thomas J. Weiler and Tzu-Chiang Yuan, \NPB318, 337 (1989); A.
Djouadi, {\it et al.}, \EPJC1, 163 (1998)[hep-ph/9701342].

\bibitem{aaa1} D.~Chang and W.~Y.~Keung,
Phys.\ Rev.\ Lett.\ {\bf 77}, 3732 (1996).

\bibitem{aaa2} E.~Keith and E.~Ma, \PRD57, 2017 (1998);
M.~A.~Perez, G.~Tavares-Velasco and J.~J.~Toscano, Int.\ J.\ Mod.\
Phys.\ A {\bf 19}, 159 (2004).

\bibitem{yuan} F.~Larios, G.~Tavares-Velasco and C.~P.~Yuan, \PRD64, 055004 (2001);
\PRD66, 075006 (2002).

\bibitem{aaa3} A. Djouadi, {\it et al.}, \EPJC10, 27 (1999) [hep-ph/9903229].

\bibitem{NMSSM}
For a detailed introduction of the NMSSM, see F.~Franke and
H.~Fraas,
Int.\ J.\ Mod.\ Phys.\ A {\bf 12} (1997) 479; for a recent review of
the NMSSM, see for example, U. Ellwanger, C. Hugonie, and A. M.
Teixeira,
arXiv: 0910.1785.

\bibitem{NMSSM-1} See, e.g.,
J.~R.~Ellis, J.~F.~Gunion, H.~E.~Haber, L.~Roszkowski and
F.~Zwirner,
Phys.\ Rev.\ D {\bf 39} (1989) 844;
%
M.~Drees,
Int.\ J.\ Mod.\ Phys.\ A {\bf 4} (1989) 3635;
%
U.~Ellwanger, M.~Rausch de Traubenberg and C.~A.~Savoy,
Phys.\ Lett.\ B {\bf 315} (1993) 331;
Nucl.\ Phys.\ B {\bf 492} (1997) 21; D.J. Miller, R. Nevzorov, P.M.
Zerwas,  \NPB681, 3 (2004).

\bibitem{nMSSM}
  C.~Panagiotakopoulos, K.~Tamvakis,
   \PLB446, 224 (1999); \PLB469, 145 (1999);
  C.~Panagiotakopoulos, A. Pilaftsis, \PRD63, 055003 (2001);
  A.~Dedes, {\it et al.}, \PRD63, 055009 (2001);
  A.~Menon, {\it et al.}, \PRD70, 035005 (2004);
  V.~Barger, {\it et al.}, \PLB630, 85 (2005).
  C.~Balazs, {\it et al.}, \JHEP0706, 066 (2007).

\bibitem{light-a-NMSSM}
  B. A. Dobrescu, K. T. Matchev, \JHEP0009, 031 (2000);
  A. Arhrib, K. Cheung, T. J. Hou, K. W. Song, hep-ph/0611211;
  \JHEP0703, 073 (2007);
  X. G. He, J. Tandean, and G. Valencia, \PRL98, 081802 (2007); \JHEP0806, 002 (2008);
  F. Domingo {\it et al}., \JHEP0901, 061 (2009);
   Gudrun Hiller, \PRD70, 034018 (2004);
   R. Dermisek, and John F. Gunion, \PRD75, 075019 (2007); \PRD79, 055014 (2009); \PRD81, 055001
   (2010);
   R. Dermisek, John F. Gunion, and B. McElrath, \PRD76, 051105 (2007);
   Z. Heng, {\it et al}., \PRD77, 095012 (2008);
    A. Belyaev {\it et al}., \PRD81, 075021 (2010);
    D.~Das and U.~Ellwanger,
  arXiv:1007.1151 [hep-ph].

\bibitem{light-a-NMSSM-1}
  S.~Andreas, O.~Lebedev, S.~Ramos-Sanchez and A.~Ringwald,
  arXiv:1005.3978 [hep-ph].

\bibitem{Gunion}
  J.~F.~Gunion,
  JHEP {\bf 0908}, 032 (2009);
  R.~Dermisek and J.~F.~Gunion,
  Phys.\ Rev.\  D {\bf 81}, 075003 (2010).

\bibitem{little-hierachy}
  R.~Dermisek and J.~F.~Gunion,
  Phys.\ Rev.\ Lett.\  {\bf 95}, 041801 (2005);
  Phys.\ Rev.\  D {\bf 73}, 111701 (2006).

\bibitem{nMSSM-cao}  J. Cao, H. E. Logan, J. M. Yang, \PRD79, 091701 (2009).

\bibitem{l2hdm-cao} J. Cao, P. Wan, L. Wu, J. M. Yang, \PRD80, 071701 (2009).

\bibitem{2HDM} J. F. Gunion and H. E. Haber, \PRD67, 075019 (2003).

\bibitem{l2hdm}
  R.~M.~Barnett, {\it et al.},
  Phys.\ Lett.\  B {\bf 136}, 191 (1984);
  R.~M.~Barnett, G.~Senjanovic and D.~Wyler,
  Phys.\ Rev.\  D {\bf 30}, 1529 (1984);
  Y.~Grossman,
  Nucl.\ Phys.\  B {\bf 426}, 355 (1994).

\bibitem{L2HDM-phenome1}
  H.~S.~Goh, L.~J.~Hall and P.~Kumar,
  JHEP {\bf 0905}, 097 (2009);
  A.~G.~Akeroyd and W.~J.~Stirling,
  Nucl.\ Phys.\  B {\bf 447}, 3 (1995);
  A.~G.~Akeroyd,
  Phys.\ Lett.\  B {\bf 377}, 95 (1996);
  H.~E.~Logan and D.~MacLennan,
  Phys.\ Rev.\  D {\bf 79}, 115022 (2009);
  M.~Aoki, {\it et al.}, arXiv:0902.4665 [hep-ph].

\bibitem{Barger} V.~Barger, P.~Langacker, H.~S.~Lee and G.~Shaughnessy,
  Phys.\ Rev.\  D {\bf 73}, 115010 (2006).

\bibitem{lsp-mass}  S. Hesselbach, {\it{et. al.}}, arXiv:0810.0511v2 [hep-ph].

\bibitem{Unitary}
A.~G.~Akeroyd, A.~Arhrib and E.~M.~Naimi,
  Phys.\ Lett.\  B {\bf 490}, 119 (2000).

\bibitem{LHWGSM} R.~Barate {\it et al.},
Phys.\ Lett.\ B {\bf 565} (2003) 61.

\bibitem{Opal3} OPAL collaboration,
Eur.\ Phys.\ J.\ C {\bf 27} (2003) 483.

\bibitem{ALEPH1}ALEPH Collaboration, \JHEP1005, 049 (2010)[arXiv:1003.0705].

\bibitem{DELPHI} DELPHI Collaboration,
Eur.\ Phys.\ J.\ C {\bf 38} (2004) 1.

\bibitem{ALEPH} D.~Buskulic, {\it et al.},
Phys.\ Lett.\ B {\bf 313} (1993) 312; G.~Abbiendi, {\it et al.},
Eur.\ Phys.\ J.\ C {\bf 27} (2003) 311.

\bibitem{Monig} K. M$\ddot{o}$nig, DELPHI 97-174 PHYS 748.

\bibitem{ffS}
J.B.~de Vivie and P.~Janot [ALEPH Collaboration],
PA13-027 contribution to the International Conference on High Energy
Physics, Warsaw, Poland, 25--31 July 1996; J.~Kurowska, O.~Grajek
and P.~Zalewski  [DELPHI Collaboration],
CERN-OPEN-99-385.

\bibitem{LEP-Report}
    [ALEPH Collaboration and DELPHI Collaboration and L3 Collaboration],
  Phys.\ Rept.\  {\bf 427}, 257 (2006).

\bibitem{Cao}  J.~Cao and J.~M.~Yang,
  JHEP {\bf 0812}, 006 (2008).

\bibitem{tau-decay}
  M.~Krawczyk and D.~Temes,
  Eur.\ Phys.\ J.\  C {\bf 44}, 435 (2005).

\bibitem{Altarelli}
  G.~Altarelli and R.~Barbieri, \PLB253, 161 (1991);
  M. E. Peskin, T. Takeuchi, \PRD46, 381 (1992).

\bibitem{PDG} C. Amsler, {\it et al.}, (Particle Data Group), \PLB667, 1 (2008).

\bibitem{2HDM-charge}
  O.~Deschamps, S.~Descotes-Genon, S.~Monteil, V.~Niess, S.~T'Jampens and V.~Tisserand,
  arXiv:0907.5135 [hep-ph].

\bibitem{Su}
  S.~Su and B.~Thomas,
  Phys.\ Rev.\ D {\bf 79}, 095014 (2009).

\bibitem{OPALchargedHiggs}
G.~Abbiendi, {\it et al.},
  Eur.\ Phys.\ J.\  C {\bf 32}, 453 (2004).

\bibitem{Davier}
  M.~Davier,  {\it et al.}, \EPJC66, 1 (2010).

\bibitem{Barr-Zee}
  K.~Cheung, {\it et al.},
  Phys.\ Rev.\  D {\bf 64}, 111301 (2001).

\bibitem{2HDM-mu2}
  K.~Cheung and O.~C.~W.~Kong,
  Phys.\ Rev.\  D {\bf 68}, 053003 (2003).

\bibitem{bsr} T. Besmer, C. Greub, T.Hurth, \NPB609, 359 (2001);
              F. Borzumati, {\it et al.}, \PRD62, 075005(2000).

\bibitem{Cao-MSSM} J.~Cao, K.~i.~Hikasa, W.~Wang, J.~M.~Yang and L.~X.~Yu,
  Phys.\ Rev.\  D {\bf 82}, 051701 (2010)
  [arXiv:1006.4811 [hep-ph]].

\bibitem{Gunion-g2}
  J.~F.~Gunion, {\it{et. al.}},
  Phys.\ Rev.\  D {\bf 73}, 015011 (2006).

\bibitem{Martin}
  S.~P.~Martin and J.~D.~Wells,
  Phys.\ Rev.\  D {\bf 64}, 035003 (2001).
\bibitem{Abdallah}
  J.~Abdallah {\it et al.},
  Eur.\ Phys.\ J.\  C {\bf 31}, 421 (2004);
 G.~Abbiendi {\it et al.},
  Eur.\ Phys.\ J.\  C {\bf 35}, 1 (2004).

\bibitem{Dunkley}
  J.~Dunkley {\it et al.}  [WMAP Collaboration],
  Astrophys.\ J.\ Suppl.\  {\bf 180}, 306 (2009)
  [arXiv:0803.0586 [astro-ph]].
\bibitem{NMSSMTools} U. Ellwanger {\it et al.}, \JHEP02, 066 (2005).

\bibitem{micrOMEGAs}
  G.~Belanger, F.~Boudjema, A.~Pukhov and A.~Semenov,
  Comput.\ Phys.\ Commun.\  {\bf 174}, 577 (2006);
  Comput.\ Phys.\ Commun.\  {\bf 176}, 367 (2007).

\bibitem{Belanger}
  G.~Belanger, F.~Boudjema, C.~Hugonie, A.~Pukhov and A.~Semenov,
  JCAP {\bf 0509}, 001 (2005).


\end{thebibliography}
\end{document}